\RequirePackage{fix-cm}
\RequirePackage[english=usenglishmax]{hyphsubst}
\documentclass[letter]{llncs}
\usepackage{graphicx}
\usepackage{microtype}
\begin{document}

\title{Designing Commercial Therapeutic Robots for Privacy Preserving Systems and Ethical Research Practices within the Home
}

\titlerunning{Designing Commercial Therapeutic Robots for Privacy Preserving Systems and Ethical Research Practices}

\author{Elaine Sedenberg         \and
        John Chuang	\and
        Deirdre Mulligan 
}


\institute{BioSENSE Lab School of Information \\
              UC Berkeley \\
              \email{\{elaine,chuang,dkm\}@ischool.berkeley.edu}           
}

\date{Received: date / Accepted: date}

\maketitle

\begin{abstract}
The migration of robots from the laboratory into sensitive home settings as commercially available therapeutic agents represents a significant transition for information privacy and ethical imperatives. We present new privacy paradigms and apply the Fair Information Practices (FIPs) to investigate concerns unique to the placement of therapeutic robots in private home contexts. We then explore the importance and utility of research ethics as operationalized by existing human subjects research frameworks to guide the consideration of therapeutic robotic users -- a step vital to the continued research and development of these platforms. Together, privacy and research ethics frameworks provide two complementary approaches to protect users and ensure responsible yet robust information sharing for technology development. We make recommendations for the implementation of these principles -- paying particular attention to specific principles that apply to vulnerable individuals (i.e., children, disabled, or elderly persons) -- to promote the adoption and continued improvement of long-term, responsible, and research-enabled robotics in private settings.
\end{abstract}

\section{Introduction}
\label{intro}
Therapeutic robots embody science fiction dreams for a better future and come with unprecedented power to analyze aspects of human behavior and health--in part through the detection of patterns in user data and continued therapeutic research and technological development. Sensors enabled by these robots can collect intimate personal data through passive monitoring and mediated interactions. For example, biosensors can capture physiological signals such as heart rate, breathing rate, and skin conductance; video cameras can recognize gestures, facial expressions, and identities of individuals and groups; microphones can detect not only voice commands and conversations, but also prosody and intonation, and difficulties with language or speech. Analysis of these multiple modal sensor data can yield surprising, and often ``category-jumping'' inferences about individuals as data are collected and mined for unrelated and unintuitive insights (e.g., deriving medical inferences from non-healthcare data) \cite{horvitz2015data}. Additionally, new trust paradigms are emerging as users connect and disclose increasing quantities and new categories of data with robots as compared to more traditional forms of ongoing monitoring like video cameras \cite{caine2012effect}. 

Therapeutic robotics represents a case where a rapidly developing technology designed to handle inherently sensitive care situations is moving quickly to market \cite{justocat2015justocat,robotcenter2015paro,robotslab2015nao}. The migration of personal robots from research laboratories and hospitals into private settings (e.g., the home or private rooms) to provide sensitive therapeutic care as commercial products exacerbates emerging privacy and ethical concerns. Combining sensors, big data analyses, and machine learning in a device that roams (figurative or literally) through intimate spaces, monitoring and analyzing bodily activities -- in this case often of the young, old, infirmed, and disabled--and intervening in daily life, therapeutic robots present the ideal conditions for a perfect privacy storm.\footnote{Similar discussions of privacy implications from smart devices in the home can be seen with regard to Mattel's ``Hello Barbie'' \cite{halzack2015privacy} and Samsung's ``Smart TV'' \cite{matyszczyk2015samsung}.} These sensors and instrumentation are generating unprecedented data that hold valuable insights about not just the therapeutic or robotic intervention but also human behavior in general. Realizing the potential therapeutic and research benefits of these robots requires the incorporation of technical and policy mechanisms to address privacy and research value ethics at the outset. 

Therapeutic robots such as NAO, Milo, and Paro are available for consumers in many countries, and though their current costs may be prohibitively high for mass consumer adoption, this is likely to change over time \cite{robotslab2015nao,milo2016,robotcenter2015paro}. As examples, both NAO and Milo both contain cameras and microphones in addition to other sensors like tactile and pressure sensors, and can establish network connections. Milo is able to record interactions and progress during sessions with an autistic child, and includes facial recognition software. Paro contains a different suite of sensors, including a microphone, and is regulated as a Class II medical device (same as automatic wheelchairs) in the United States. This classification does not regulate the uses and storage of information collected, but focuses mostly on labeling and ability to recall devices should a safety defect be found \cite{fda2016medical}.

As the consumer market for robots with therapeutic applications develops, user data may be available not only to healthcare providers, therapists, and researchers as before, but also to the companies who manufacture these robotic devices and data platforms. Each of these actors in the U.S. are subject to different legal and regulatory regimes when deploying these systems, which means there will be many different design and policy requirements that come with consequential privacy and data use outcomes for the consumer. Further, unless attention is paid to the research potential of these robotic devices and platforms, researchers (in both private industry, medicine, and academia) may experience difficulty using the robots or the data generated due to ethical objections.

There is a window of opportunity to incorporate privacy-preserving and ethically responsible research frameworks into the overall design of commercial therapeutic robotic applications for use in the home -- whether they are care functions, physical or emotional therapeutics, at-home medical aides, or tele-operated care. Doing so would ensure that these systems are ``research ready.''  Given the valuable research insights that might be gained by user and data studies on them, the failure to attend to these issues would be a tremendous loss to the research community. 

Our analysis focuses on U.S. law and related policies because they are still under development and may benefit from recommendations and market guidance. Though the principles discussed here may apply more broadly to any robotic device or even other ubiquitous computing objects within the home (or perhaps in other sensitive settings), therapeutic robots provide a particularly salient vehicle for highlighting the opportunities and risks for users and researchers, and the need for guidance at this critical moment of commercial readiness. 

First we discuss related literature and highlight the novel contributions of our analysis and design suggestions. We then examine the privacy concerns of users of commercial therapeutic robots within the home and present new privacy paradigms alongside the Fair Information Practices (FIPs) to investigate concerns unique to this placement. We next explore the complementary utility and importance of research ethics (as operationalized by existing human subject research frameworks) for the consideration of therapeutic robotic users -- a step vital to the continued research and development of these platforms. Together, privacy and research ethics frameworks can protect users and ensure responsible yet robust information sharing to support research and further technology development within and outside academia. We make recommendations for the implementation of these principles -- paying particular attention to specific principles that apply to vulnerable individuals (i.e., children, disabled, or elderly persons) who are likely users of these devices -- to promote the adoption of these frameworks for long-term, responsible, and research-ready robotics in private settings. We intend this discussion and recommendations to protect users while enabling robust research and development activities, in part through the availability of responsible data sharing, that continue the improvement and social impact of these robotic platforms.

\section{Relevant Background and New Paradigms}
\label{sec:background}

Prior work explores the ethical, privacy, and security implications of robotic systems and the integration of ubiquitous computing systems in the home, which come with unique vulnerability and privacy challenges for the end user  \cite{edwards2001home,hong2004architecture,sung2007my}. For example, Lisovich et al. examined what behavior inferences could be made about users from smartgrid utility data \cite{lisovich2010inferring}. Denning et al. introduced security and privacy design questions that begins to explore some of the long-term privacy aspects of at-home robotic use \cite{denning2009spotlight}. Legal analysis by Lerner and Mulligan examined how data collected by smart devices (specifically smart meters) may interface with existing American privacy laws specific to the home \cite{lerner2008taking}.  Kaminski analyzed the legal aspects of records generated by at-home robotics, and Calo explored the unique privacy risks flowing from the tendency to anthropomorphize robots. \cite{kaminski2015robots,calo2010robots}. These unique privacy risks include intrusions on self-reflection and increased collection of information about sensitive traits and behaviors through sensors and programmable settings. Some literature has explored the importance, utility, and potential barriers to adoption of therapeutic robots in the home, for uses like aging in place, receiving at-home healthcare, and autism therapy \cite{robinson2014role,alaiad2014determinants,cao2015probolino,khosla2015socially}. Though there are independent discussions of informational privacy implications for data collected in the home and need for therapeutic robots in these private settings, prior work does not analyze these issues simultaneously. 

Past studies have shown that there is a privacy-utility tradeoff that can be balanced, especially with regard to additional sensors like visual feeds, which was demonstrated by Butler et al. in the case of teleoperated robots within the home \cite{butler2015privacy}. Further, the presentation of sensors or extra data streams results in material differences in how users modulate their privacy preserving behaviors, as seen in a recent study of older adults and visual monitoring \cite{caine2012effect}.

There is a rich body of research that tackles the myriad of ethical dilemma complexities of human-robot interactions \cite{lin2011robot}. For instance, Decker reflects on the ethics of robots as a care replacement and examines the implications of autonomous robots with the capacity to learn (and thus make their own decisions independent of the manufacturer) \cite{decker2008caregiving}. Sharkey and Sharkey discuss the ethical concerns raised when robots replace human caregivers for the elderly, and attempt to balance the benefits and costs of these robotic interventions \cite{sharkey2012granny}. These robotic ethics papers tackle difficult questions regarding when and how to institute care robots or therapeutic interventions mediated by robots. In this paper, we focus our attention on ethical research practices for therapeutic robots, driven by the fact that the users of these robots may be considered human subjects in research studies, even when the robots are purchased by consumers and operating in private home settings. Though the existing body of work on robot care ethics may overlap with many of the aims of research ethics by seeking the same or complementary values to protect the end user, there are unique risk and benefit tradeoffs inherent to human subject research. These tensions are particularly salient considering the continued research and development of commercial therapeutic robots. Attending to these issues is particularly urgent since sites where research occurs and individuals involved expand beyond academia, thus outside of current norms and practices for human subject research.

While most current studies of the general use and effectiveness of therapeutic robots are conducted within academic contexts, future studies can and will be performed in the field with actual users by private companies and/or their collaborators. However, once therapeutic robots leave the laboratory and clinical setting, they will no longer be regulated for ethical oversight\footnote{An exception would be if a private company receives federal grant money to fund a study, but even academic collaborations often do not trigger Institutional Review Board (IRB) oversight. Some companies have private ethical review practices or private IRBs, but these actions are not regulated or standardized.} \cite{us2009code}. Riek and Howard point out this gap in ethical oversight and call for human-robot interaction (HRI) practitioners to develop a code of ethics and propose a set of principles to grapple with the implications of research, development, and marketing techniques in the private sector \cite{riek2014code}. A similar gap in ethical obligations has been noted by those concerned with basic scientific research occurring on social network sites \cite{grimmelmann2015law}. The addition of research ethics frameworks complements the assertions from the paper. Private sector researchers may perform natural experimentation observations and algorithmic tailoring with methods such as A/B testing or multi-armed bandit optimization \cite{scott2010modern} to improve these systems. Data-driven research and algorithmic adaptations to our individual differences can lead to new knowledge about populations as well as particular users, which underscores the imperative that designers have to consider unintended consequences for privacy, as well as long-term risks and benefits. These opportunities come with significant ethical considerations for the users--particularly for vulnerable\footnote{The Belmont Report refers to vulnerable populations as those that ``either have limited capacities to consent, have subordinate relationships to the investigator or his institutions...or -- by virtue of other aspects of their life -- are especially vulnerable.'' The Report goes on to specify that those with ``limited capacities to consent'' include children, fetuses, prisoners, mentally institutionalized, those under the influences of addiction, or those otherwise vulnerable as a consequence of their life situations (for example, those legally enfranchised to grant consent but are in reality incapable of sufficient comprehension, persons with prolonged illness) \cite{national1978belmontAppendix}.}  robotic therapeutic users--that must be balanced to fully realize the potential benefit of these technologies. The integration of research ethics frameworks into the information systems and deployment of therapeutic robots operationalizes core values (respect for persons, beneficence, and justice) and further protects robotics users. These frameworks unlock research opportunities on these platforms as well as data sharing for research purposes.

\section{Personalized Privacy for Personalized Robots: Grappling with Data Intimacy and Permanency}
\label{sec:personalized}

Privacy is a dynamic concept that is context dependent and evolves as political and technical features develop in modern society \cite{moor2006using}. Scholars have proposed four distinct kinds of privacy: physical, decisional, psychological/mental, and informational. The purchase and use by individuals signals a decision to bring the therapeutic robot into the home, so we will not focus on physical privacy and intrusion but is worth additional analysis in future work. Similarly, decisional privacy or interference with one's choices we feel deserves additional attention, particularly with regard to  ``nudging'' behavioral adaptations in therapeutic applications, but is out of scope for this paper. We instead focus our analysis specifically upon informational privacy, but touch briefly on psychological/mental privacy as evoked by growing use of physiological sensors and emotional facial recognition later in the section \cite{tavani2008informational}. 
Personal information relevant to the concept of informational privacy includes data collected by sensors, generated by a users' interactions with a device, or communication data as argued by Johnson and Nissenbaum \cite{johnson1995computers}. Information privacy impact can be analyzed in terms of the amount of information collected, speed at which it may be exchanged, duration of retention, intended or possible recipients, and kind of information collected \cite{tavani2008informational}.

Calo breaks the privacy impact of robots into three broad categories that are each relevant to information privacy: direct surveillance via sensors to new actors (e.g., private corporations as well as individuals and governments), increased access (nature of co-placement in the home opens creates unprecedented technological windows), and social meaning (new trust paradigms impact the way information is shared with technology via robot interactions) \cite{calo2010robots}. This creation of new data streams and windows is further underscored by the sensitivity of the placement and use of therapeutic applications.

In the remainder of this section, we examine the potential for informational privacy harms specific to the capabilities of therapeutic robots, provide privacy vignettes that illustrate these tensions and tradeoffs, then apply privacy frameworks to inform actionable design recommendations that embrace privacy-by-design principles.

\subsection{Aspects of Informational Privacy Specific to Therapeutic Robots}

Many therapeutic robotic devices are designed for long-term usage and placement with an individual within sensitive, and often intimate, settings \cite{jibo2014meet}. These robots and their underlying data may cover a significant portion of an individual's maturation (e.g., autism therapy) or end of life care (e.g., elderly companions), which cover sensitive developmental timespans and hold the potential to collect very large amounts of information even with very few sensors. Further, the data generated by the robots may last much longer in repositories than the condition underlying robot delivered therapy. Privacy and data management policies and practices are often not proportionally designed to accommodate these extended timescales, sensitive contexts, and potential permanency of such high-volume data. The timeline for retention of data is rarely specified by private companies in the information sector, and given the context of therapeutic interventions may be more material to users within these use cases. 

Health privacy laws do not extend to the use of private devices and in-home collection possible using privately purchased therapeutic devices. Outside of a strict healthcare context, many of these robotic devices may embrace market norms of voracious data collection and reuse--not appropriately suited for a therapeutic healthcare application. Data collected may disclose detailed aspects about individuals' health or mental status, disabilities, and vulnerabilities--in addition to particulars regarding treatment regimes, compliance, and subjects' responsiveness to particular therapies. Inside the U.S., the Health Insurance Portability and Accountability Act (HIPAA) applies only to covered entities including: health plans, health care clearinghouses, healthcare providers who transmit any health information in electronic form. \cite{hhs2006HIPAA} Even though some therapeutic robots may be classified as a medical device as previously mentioned, the level of classification  does not provide restrictions on information flows. Therapeutic robots sold on the consumer market would come under the regulatory purview of the Federal Trade Commission (FTC) for unfair and deceptive practices -- regulatory oversight that has been used to police privacy--but would be reactionary instead of proactive guidance \cite{ftc2006code}. 

Co-location of devices in private home settings will implicate more individuals than the primary robot user or recipient of therapeutic applications. For instance, simple audio and visual data may capture information about family members, other minors like siblings, or friends and visitors who may come to home environments unaware of the potential for mass data collection through therapeutic robotic agents. In the United States, like many places around the world, the home is considered a highly private context in which individuals conduct themselves with a higher expectation of privacy \cite{nissenbaum2004privacy}. Unlike hospital or classroom settings where the context of the environment remains mostly static (i.e., hospitals are for treatment and care; classrooms are for educational contexts), the home is a complex and dynamic environment that may be both a social context when friends are over, place of care for sick individuals, or a place of private conversations and activities. The variety of activities in the home and the heightened privacy expectations should be taken into account.

Many therapeutic robots are emerging on the market as Wi-Fi and Bluetooth enabled so they can receive commands, push data to the cloud, and communicate with other devices. Though the ability to connect robots to intra- and extra-sensors is only beginning, the potential to work with wearable physiological sensors and other Internet of Things (IoT) devices within the home greatly expand the capacity of these devices to pool and compare data in unprecedented ways. This interconnectedness may lead to novel combinations of data and generate new (perhaps unintended) knowledge about individuals, which creates the potential for new informational privacy harms in terms of not only the new amount of information collected, but also the speed and variability at which it may be exchanged \cite{tavani2008informational}. In addition to potential informational privacy harms, images of users can be used to infer psychological and mental states which violates an entirely new category privacy interests that may be unexpected and poorly understood by the user--especially since the inferential possibilities of these analyses are not yet fully realized or comprehended by even experts. Though this type of privacy harm is not well examined with respect human-robot interaction, we believe it is immensely important given the spectrum of therapeutic functions relating to cognitive or emotional impairments and should be specifically examined in future work.

\subsection{Privacy Vignettes}

To further demonstrate how therapeutic robots are in a unique position to frustrate current privacy norms due to the continued development of these technologies, we present the following vignettes. The capabilities of the therapeutic care reflect direct technical capabilities of robots entering the commercial marketplace or represent minor developments that can reasonably be expected given contemporary IoT devices currently on the market. We will then explore two applicable privacy frameworks to constructively address these issues.

\textbf{Extended Timescales for Data:} A 7-year-old child with a learning disability receives at-home supplemental instruction with a new learning robot his parents purchased to help improve is reading and speech abilities. The child uses the robot for 3 years before discontinuing use because the developmental delay is no longer noticeable. The data and records however persist with the learning robot manufacturer, and when a larger company acquires the smaller device manufacturer, the data about the now 30-year-old individual is merged with a growing corporate digital dossier.  The contents of this new database now includes details about the individual's early reading and speech development, which becomes meaningful when new predictive algorithms begin to market the latest developmental technology to the individual's new baby.

\textbf{No longer healthcare data:} A 60-year-old woman with early onset Alzheimer's disease takes part in a medical research study that uses a therapeutic assistive robot to assist with daily memory tasks. After the study conducted under the supervision of her doctor concludes, the woman purchases the therapeutic assistive robot to use at home in hopes that it would slow the early progression of the disease. Now that she is using the robot at home and not under the guidance of her doctor, her data is no longer protected by U.S. federal privacy law and her medical condition information (relayed through her purchase of the robot and subsequent collection of user data and degenerating cognitive abilities) is sold to data brokers. The data is used to deny the woman's desire to switch life insurance in preparation for her family because the use of such information to make insurance determinations is not prohibited in the U.S.

\textbf{Unwilling participants:} A couple purchases a therapeutic robot to assist their autistic child in learning social cues at home. The robot records interactions with the child so that it may learn and adapt to the specific needs within the autism spectrum. The robot contains video and audio capabilities, and connects to Wi-Fi in order to store data in a cloud-based server. The child has a friend from class that often comes over, and they frequently play with the robot together. Neither the friend from class or his parents are aware a robot is frequently recording their child's interactions and storing this information in a server they cannot access.

\textbf{Home as a fluid context:}  An elderly man is receiving terminal end of life care within his own home, and was given an emotionally assistive robot to promote stress reduction. The robot has basic machine learning features that help it customize sessions to the owner. Late one night after the elderly man has gone to bed, his children sit in the living room discussing what they are going to do to clean out the house after he passes. One of the children picks up the emotionally assistive robot, and accidentally activates it as she strokes the pet-like robot. Part of their conversation is recorded in the home, and accidentally is brought up the next day when the emotionally assistive robot prompts the elderly man about cleaning out his house. His daughter is embarrassed and uncomfortable because she and her siblings thought the conversation was private in their own home, and was discussed intentionally while he was asleep.

\textbf{Care without a professional intermediary:} A woman buys herself a therapeutic robot at the urging of her therapist to help her cope with a recent bout of depression. She takes a break from her weekly therapy sessions at the encouragement of her therapist, but continues to use the robot which gives her feedback on her progress based on their interactions. Though the woman continues to improve and feel better about her mental state and overall wellness, the robot through emotional recognition provides feedback that she is not improving. Without her care intermediary (her therapist), the woman does not understand this feedback is only based on facial emotional queues (which have an associated error rate) and not other psychological metrics. She is overwhelmed by the supposed lack of progress. 

These vignettes are meant to illustrate privacy issues presented in situations specific to the places, types, and unexpected flows of information.

\subsection{Application of Privacy Frameworks to Elucidate Principles}

In order to address the issues and tensions highlighted in the vignettes, we explore two frameworks--contextual integrity and the Fair Information Practices--for considering the new privacy paradigms introduced by the co-location of therapeutic robots in the home. 
Nissenbaum's theory of privacy as ``contextual integrity'' (CI) connects privacy protections to the informational norms of specific contexts \cite{nissenbaum2004privacy}. Nissenbaum's central claim is that ``finely calibrated systems of social norms, or rules, govern the flow of personal information in distinct social contexts.''  \cite{nissenbaum2009privacy} Understanding and respecting privacy requires context specific investigation, and norms developed in one sphere are often inapplicable to another. Moreover, ``distributing social goods (such as the protection of personal information) of one sphere according to criteria of another constitutes injustice.''  \cite{nissenbaum2004privacy} Nissenbaum's theory foregrounds the \textit{practice} of privacy, and urges attention to its essential embeddedness, rather than essential meaning. 
Working with CI directs attention to specific spheres of social life and the informational norms that support the activities within them. Within the context, CI directs us to take stock of how specific types of information about individuals in particular roles are governed: What information flows between which parties, and under what conditions? 

CI offers a useful framework for evaluating the privacy issues posed by the introduction of therapeutic robots into personal places. When moving care from hospital setting to home -- what are the norms? Robots that may not be classified as medical devices but consumer devices -- what are the norms? Providing care to vulnerable individuals and learning about effectiveness to improve outcomes through research -- what are the norms? Data flows leaving the home (a space traditionally considered private and strongly protected by law) -- what are the norms? In part, CI helps to uncover these information exchanges that may interface with existing or need for developing norms.

The internationally recognized Fair Information Practices (FIPs) are the dominant framework for protecting information privacy \cite{gellman2014fair}. FIPs embody a rights-based framework of informational self-determination.  In contrast to CI, FIPs reflect a liberal and essentialist construction of privacy that seeks to support ``the claim[s] of individuals . . . to determine for themselves when, how, and to what extent information about them is communicated to others,'' \cite{westin1968privacy} through the imposition of processes that support the individual's agency over the flow of personal information, and establish obligations on data holders across all contexts. FIPs requires the adoption of policies and mechanisms through which ``individuals can assert their own privacy interests and claims if they so wish,'' allowing them to define ``the content of privacy rights and interests.'' \cite{bennett2006governance} Privacy in this framework demands processes that empower individuals to make fully informed decisions about their personal information. CI foregrounds context and the goals, activities and relationships within them. FIPs, in contrast, foregrounds essential rights and obligations that largely transcend such contextual concerns.

It is vital that robotic devices, regardless of therapeutic application, be considered as a whole and inclusive of their information systems--tracing data and information flows from collection via sensor or manual inputs, transmission, storage, and access. By considering the data and information flows as part of the robotic system, it is possible to take a more specific analysis of the privacy implications and make specific recommendations to improve user controls and minimize risks.

The original 1973 FIPs version points to core principles including: 

-Transparency (no secret systems) 

-Access (to individuals' records and their uses)

-Privacy Controls (ability to prevent information about oneself from purposes without consent)

-Integrity (ability to correct or amend)

-Data Use Protections (prevent data misuse)

Transparency, as is implied, emphasizes open practices that assist users in understanding information systems collecting data about them. But transparency also articulates the need to ensure there are ``no secret systems'' or information flows that the user might be unaware of. Similarly, a user should be able to access information records collected about them and understand how these data are being used. For example, if an audio sensor is embedded and records human-robot conversations for natural language processing to ``learn'' names or pick up on commands, the user might be unaware that this also entails storing audio records for unknown amounts of time, or may not understand that the microphone also picks up ambient conversation while listening for its name. The collection and storage of information (including data), even if it is not stored for an extended amount of time, should be clearly explained and made transparent and accessible to the user under these principles.

Once a user understands the information flows within a system, they are better able to make privacy decisions and should be able to set controls on these information flows within different contexts. These controls could include where data are stored (locally or in the cloud), how long they are kept, which sensors are enabled, how information is used, etc. In the case of therapeutic robots, this may be especially important as the variety and sensitivity of home activities may present a need to change preferences fluidly over the use application of the robotic device. 

Integrity of data, especially as it is used for decision-making, algorithmic adaptations, and record-keeping is vitally important and in accordance with this principle individuals should be given recourse to correct or amend information logged about themselves. For instance, if a visitor interacts with the robot but the data is captured and stored as part of the therapeutic subject's files it could interfere with appropriate treatment, and it could expose sensitive information about the visitor. A design that limits the mis-association of different individuals data (either by not collecting other peoples data or segregating it) is essential to support this principle. System designers should put mechanisms into place that keep data secure and prevent unauthorized access and use. For instance, controls should be set on who could view data generated by a robotic device, or security protocols should be automatically updated to protect the integrity, availability, and confidentiality of the device.

\section{Ethical Frameworks to Enable Robust Research and Data Sharing Practices within the Private Sector}
\label{sec:ethical}

Private companies are likely to have engaged in extensive research to develop these personalized robots for all use applications (including but not limited to therapeutic applications), and will continue to be interested in performing research using the deployed robots as platforms in order to gain new insights into user interactions and how robots may adapt to users' individual differences and preferences. More importantly, these new devices and platforms offer an unparalleled opportunity to advance scientific research by supporting research in the wild in a relatively noninvasive and inexpensive manner. Not all of these research activities will occur within academic, clinical, or educational settings that are equipped (and regulated) to protect subjects or users. Thus, it is important to develop ethical frameworks for responsible research and development (R\&D) practices involving users and their data that focuses on an ethical application of principles that expands beyond the current discourse in robot ethics. We will first explore why the area of research ethics is important to the therapeutic robot community, then present principles that operationalize human subject research values as a foundation for our recommendations in the next section.

Research ethics seeks to balance the risks and benefits inherent to research activities with regard to the individual, while simultaneously considering the potential externalities to society as a whole. Many of the principles and values research ethics aims to protect (i.e., respect for persons, beneficence, and justice) may be complementary to those present in other areas of robot care ethics (e.g., user autonomy, integrity, respect for the human condition), but specifically take into account the need to consider circumstances where tradeoffs may be necessary to incrementally benefit the human subject user during therapy (and more likely eventually society as a whole). For example, minimizing all risk to preserve value is sufficient for robot care ethics in general, but for research activities some risk (e.g., diminished privacy, risk of data breach, discovery of unintended knowledge, potential for a less optimal therapy, minor discomfort to patient during research, burden on user time, etc.) may be necessary in order to improve the quality of therapy or assist in scientific discovery. Without an ethical framework to determine acceptable activities, procedures, and risk/benefits tradeoffs, we risk infringing on core values that could otherwise be minimized. Without any guidelines for industry on how to balance value tensions in a way that preserves core ethical principles while still allowing research activities, we could lose out completely on potential improvements to robotic therapies. These improvements not only benefit greater society, but the human subject end user as well. 

In the U.S., ethical oversight boards regulate only medical and federally funded human subject research. As a result, much research occurring in or funded by the private sector falls outside legally driven ethical frameworks. This is problematic on two counts: 

1) Robots put on the market by private industry will undoubtedly continue to be improved through internal R\&D practices, which will likely stem from data and interactions collected from end users. The internal research necessary to make these improvements and develop robust therapeutic applications will hinge upon both broader behavioral and individualized human study. Though there is current debate over what constitutes an experiment for generalizable knowledge and what practices are merely intended for product development, respecting the interests of users--particularly children, disabled, elderly--and ensuring these devices and platforms are viewed as appropriate venues for research council adoption of ethical frameworks generally.

2) Private companies deploying therapeutic robots will come into a treasure trove of valuable user data that would benefit many academic and healthcare or education researchers who work on improving human robot interaction, therapeutic efficacy, user privacy, etc. Recent controversies over data collection and collaboration between industry and academia, like the Facebook contagion study, highlight the uncertain and tenuous ethical grounds when platforms are shared for mixed commercial and academic research purposes \cite{kramer2014experimental}. Given the powerful public good potential of these data collected by private companies on therapeutic robot platforms, these systems should be designed to support sharing for research purposes. Incorporating human subjects research principles into the information system design enables responsible transfer between industry and academia. If elements like informed consent as discussed below are incorporated into the initial design and deployment of therapeutic robots, private companies and users will benefit from researcher access aimed at improving therapeutic outcomes. But researcher access within the United States may be restrained if basic human subjects requirements are not fulfilled at the onset of data collection given evolving norms within human subjects regulations.


\subsection{Development and Relevance of Historical Human Subjects Frameworks}

When human subject experimentation regulations were developed in the United States, ethicists and practitioners sought to operationalize core principles (respect for persons, beneficence, and justice) through the development of frameworks via the Belmont Report (and later expanded by the Menlo Report), which led to eventual policy and legislation \cite{national1979belmont,us2009code,dittrich2011menlo}. Outside of strict regulatory compliance, these frameworks and policies provide useful implementation guidance on integrating ethical values into technology and data system design. Thus, we apply the canonical ethical frameworks from the Belmont and Menlo reports to the design and implementation of therapeutic robots, and discuss opportunities for optimizing their application to maximize benefits and minimize risk while enabling robust human-centered research. These frameworks provide special considerations for vulnerable persons, and highlight the importance of scientific research and the production of substantial societal benefits through innovations.

Belmont presents three ethical principles:

-Respect for persons (individuals should be treated as autonomous agents)

-Beneficence (obligation to maximize benefits and minimize harm)

-Justice [distribute benefits of research fairly and careful selection of research subjects (e.g., avoid only selecting from underserved groups which places undue burden on part of the population)]

Informed consent is one application of these principles, and particularly embraces the principle of respect for persons since the process of informed consent allows individuals to opt-in or out of the research activity, thus exercising autonomy. Consent as operationalized within the report consists of three core elements: information, comprehension, and voluntariness. Information conveyed through informed consent should contain details regarding the research procedure and purposes, risks and benefits, alternative procedures where applicable, and a statement allowing the subject to ask questions or withdraw at any time. Comprehension is achieved through presenting information in a manner and context that promotes understanding (i.e., clear and in a way that allows questions) and the Belmont Commission suggests that it may be suitable to give an oral or written test for comprehension. 

Voluntariness is the final component of informed consent, and protects against coercion or unjustifiable pressure. Nuremberg Code was the first set of ethical guidelines established to govern the use of human subjects in research, and heavily emphasized voluntariness after Nazi prisoners were forced into scientific experiments \cite{code1949nuremberg}. Voluntariness may be implemented by only opting individuals into research with their expressed consent to participate, and informed consent containing the elements described above is one mechanism to promote voluntariness. 

Further attention may be necessary to look at the diversity of groups included in studies to make sure that no one subpopulation bears an undue burden of research without standing to reap most of the benefits. The recognition of the importance of justice when selecting and using groups of people for research is important but not easily formulated. Research that advances the interests of those being studied is preferred more than those of the general population or particularly of another subset of individuals.

When assessing risk and benefits, Belmont provides a framework of considerations including balancing both the probability and severity or magnitude of the potential risks. Risks to be assessed include psychological, physical, legal, social, and economic. Benefits should include those to the individual and society as a whole, but place special emphasis on the benefits that may be expected by the individual person. This means that if any risk is likely or a high magnitude risk is placed upon an individual, they should stand to directly benefit from the technical intervention or else the balance of these two considerations fails.

In 2012, the Menlo Report built off of the Belmont principles specifically for Information and Communication Technologies (ICT) research, and added an additional principle: Respect for Law and Public Interest. This principle prompts engagement in legal due diligence (compliance with existing data or communications laws), be transparent in methods and findings, and be accountable for actions [31]. This principle promotes transparency and encourages designer accountability for interventions, which is appropriate within the robotics context. As in the general ICT environment, therapeutic robotics raise a range of legal issues (as partially noted in the related work section). Given the overlap and connection between legality and ethics, close attention to existing law is important.

Special ethical considerations for vulnerable persons are particularly relevant to therapeutic robotics, and some guidance is offered through research ethics and existing laws. Robots aimed at working with children, handicapped individuals, or the elderly may face increased regulations aimed to protect vulnerable individuals. For instance, in the United States the Children's Online Privacy Protection Act (COPPA) regulates the collection of data from children under the age of 13 by commercial websites and online services (including mobile apps), and includes provisions governing parental consent, data-use, and retention \cite{16cfr312}. While COPPA will have limited if any direct application to therapeutic robots, the FTC has used its general ability to police unfair and deceptive practices to address the heightened privacy risks to children presented by information and communication technologies. It will no doubt do so in this area as well.

Enabling appropriate data collection and interactions with vulnerable individuals provides substantial societal benefit by enabling research on children with autism, assisting elderly adults, and other conditions that may require research attention \cite{broekens2009assistive,liu2007affect}. The national commission behind the Belmont Report also considered special research cases on children and people with mental disorders as individuals who have reduced autonomy, and examined considerations of comprehension and ability to grant informed consent \cite{jonsen1978research,national1978research}. The commission determined that so long as undue risk without individual expected benefits was not placed on vulnerable people, research should be permitted since research is necessary to develop new treatments and insight that uniquely affect these populations. These reports suggest that a third party (i.e., a parent or guardian) be chosen to represent vulnerable individuals' interests when giving informed consent while also allowing the individual themselves an opportunity to choose their participation to the extent they are able.

The ability to respect and preserve the autonomy of individuals, balance risks and benefits, and aim for equally serving (and not unjustly burdening a group of people) provides a framework for creating ethical research practices that may apply to the research and development of therapeutic robotics technology.

\section{Implementation Recommendations}
\label{sec:implementation}

Given the unique challenges of commercialized therapeutic robots discussed within this paper, we present a series of design and implementation recommendations to optimize the protection of users and the realization of research potential of these emerging devices and platforms. Privacy and research ethics can be implemented through system design early on in the technology deployment so that responsible information practices, including sharing, may occur. 

\textbf{Data Access and Review:} There are many design options that would enhance individuals' ability to understand and manage their information collected by therapeutic robots -- including access to audio, video, or any other data collected by the platform. The process of granting access to users additionally increases transparency of the information systems involved with the robotic platform and thus prevents the existence of secret systems referred to in the FIPs. 

Alongside access and transparency, FIPs points to the integrity of data through the ability of subjects to review and amend their records for inaccurate or incomplete data. In addition to this ability to correct, the ability to retroactively omit information would also help users maintain the integrity between shifting context and shifting data. Users should be given the option to amend incorrect data, since algorithmic misinterpretations over recorded actions could result in frustrating or privacy-diminishing suggestions made by the robot. Ideally, retroactive control of data privacy settings would allow users to change archived data in nuanced ways instead of allowing for a binary choice of all or nothing deletion controls. Data integrity is of paramount importance to research activities, so if a user sees inaccuracies in the data record during review, there should be a clear process to submit a correction. 

Since therapeutic robots may be used in contexts where users may have diminished autonomy, access should be granted to guardians or caregivers. However, preference should be given to the individual receiving care to the extent possible, or in cases where maturity or recovery are possible, this access should be given in the future so that users with once diminished autonomy may regain control of their data legacy.

Special consideration should be given from the designers of therapeutic systems to interrogate who, in the cases of users with diminished autonomy, should be granted access. For instance, should parents be given access to all of a teenager's behavioral data or only relevant data to a therapeutic intervention? Should multiple children have access to their elderly parents' data or just one selected primary caregiver? Should data from a healthcare setting be shared with family first before an external care provider so that certain personal privacy may be protected? There is no right answer and should be left up to the system designers, or with assistance from the designers, handed off to the end users. Determining access and review provisions to protect therapeutic robot users with diminished autonomy will certainly not be a one-size-fits-all model, but is an element essential to designing a privacy-preserving, ethical system.

\textbf{Presentation of Privacy Policies and User Consent:} Consent for participation in research as well as data use, though two distinct concepts that may require additional unique descriptive elements (e.g., risks and benefits for research studies), may functionally go hand in hand. Personal robotics (especially those used for therapeutic contexts) should take advantage of the diverse functionality of the robotic platform and offer consent and notification through multiple modalities. For robotics with a limited screen interface, text could be sent to each user on a separate device using email, or agreement could be reached through an auditory explanation and visual confirmation of the users' understanding and consent using the robotic platform. 

Instead of limiting consent to binary decision, occurring through paper at specific moments in time, the robotics industry should embrace dynamic consent models that allow for the user to select more nuanced participation choices (e.g., use all of my data, or only use my data for certain types of research), protocol updates over time, and the ability to change decisions or be notified of scientific findings over time \cite{stopczynski2014privacy}. Features of many therapeutic robots would also allow for improved informational content within the consenting process and testing for comprehension prior to accepting any agreement through user interactions with the robot itself.  When initial consent is put into place, it should be designed with the intention of future contact between the research platform and the data subject, even if the data subject gets rid of the research-enabled device. These dynamic and flexible consenting procedures may help improve the ability of individuals with diminished autonomy to maintain their understanding where possible (e.g., through the use of visual or auditory presentation of simple concepts), but reinforce the autonomy of the consenting individual overall by improving accessibility and potentially the informative capacity.

Better consent not only improves upon the user's ability to determine privacy permissions, but also operationalizes the value of respect for persons as illustrated in the Belmont Report. Improving consent aligns with research ethics so that as private companies use data for development, collaborate, and seek to operationalize core ethical values (i.e., respect for persons). Implementing these consenting procedures creates the potential for more robust and responsible information sharing, and could incentivize collaborations and further research -- points we believe will be vital to the continued improvement of therapeutic robotics platforms.

Regarding the timing of consent: it should be obtained from individuals prior to using data in a research study, and if the robot is research-enabled from the start consent should be obtained immediately. In some cases, timing of consent must be offset to avoid bias within a study. It is preferable to garner consent many months in advance so that subjects may approve but forget about the planned research activity, but if this is absolutely not possible researchers should not use any data collected until consent has retroactively been obtained. If the individual does not consent ex poste facto, data must be destroyed.

As a final point to consent, individuals should be notified of scientific findings derived from their participation, so that they receive positive feedback and may possibly benefit from the research produced. This added potential for benefit by the end user helps mitigate risk tradeoffs involved with aggregate level study of their data, and reinforces principles of beneficence from the Belmont Report.

\textbf{General Privacy Controls:} In addition to permissions and consent regarding information flows and data practices, users should also be given choices that allow for the direction and control of their robotic information system. For example, choosing whether data will be stored locally or archived in a centralized repository (and for how long) gives users the autonomy to control how at least some of their information is accessed or used. Related to the concept of dynamic informed consent, designers should anticipate that user preferences for general privacy controls will evolve and change over time. There also may be circumstances where even after controls and rules over a system have been set, situations arise where users should be reminded of the implication of their choice similar to how smart phones will remind users GPS location tracking is still active. For instance, devices should always make image capture clear through the use of a light or sound, and be designed to remind the users that recording is still active if a camera detects flesh more flesh tones (implying someone might be accidentally naked in front of the camera) or if the microphone detects arguing (indicating a sensitive conversation not relevant to the therapeutic application may be occurring). There may be cases where multiple users need to set multiple preferences--a feature that could be enabled by facial recognition possible in many therapeutic robots -- but given our scope of at home personal uses this may not be a primary issue. Also as a rudimentary solution that is not always viable with the introduction of ambient sensors, these robotic devices should also be easy to shut off as a master control so users have maximum control over timing of data collection. 

Though adding these design considerations and controls adds some extra development costs, it would improve usability and potentially user satisfaction with the product. It would also ensure that researchers can use the platform and data derived from it in a manner consistent with ethical obligations and expectations. Further, these controls and access measures will decrease potential liability for companies stemming from unfair or deceptive claims by users or regulatory bodies. 

\textbf{Awareness of existing laws and potential data use:} Under current U.S. law, personal data and communications stored by robotic devices will be accessible to government agents under different standards and processes depending upon how they are legally classified. Data collected by robots but stored on company servers would fall outside the Fourth Amendment due to the Third Party Doctrine, and be available to the government without a warrant and depending upon the service being provided, context of use, and entity, potentially without other legal process \cite{thompson2014fourth}. In contrast, data stored on a therapeutic robot owned by the user would in most instances require a warrant for government access. Different countries yield different protections, and such variances are often unanticipated by users and practitioners. These risks of potential data access should be disclosed to the user since the average person may not be aware of how these data (which may hold particularly revealing content) may be accessed by law. By notifying users of legal implications of service for personal data with existing laws, robotic companies would comply with guidance from the Menlo Report by dually respecting the law and public interest. It is likely most companies will be aware of these laws internally, and providing some extra notification to users will come at a low cost to the manufacturer. 

\textbf{Design for Responsible Data Sharing:} Therapeutic robotic designers should consider building privacy-preserving data sharing mechanisms into research-enabled robots, so data that has been ethically collected may benefit a wide spectrum of researchers and topics. Since therapies employing the use of robots are rapidly evolving and under constant improvement, robotic devices sold by the private sector would benefit from policies that enable responsible information sharing with outside researchers or partners. Research using data from these robotic platforms, if done in an ethically aware manner, could improve the commercial device via otherwise unrealized research partnerships. Many proposals for open Personal Data Stores (PDS) have been made in the past, which allow for the encrypted storage of personal data and remote access to datasets \cite{anciaux2013trusted,de2012trusted}. By maintaining control of data, individual researchers only gain access to large-scale analysis of data and not underlying or irrelevant information. Many designs for PDS also contain provisions for changing permissions or revoking data access.

There are many proposals for PDS designs, including the open PDS model which allows for the encrypted storage of data where the user allows researchers to remotely run their research code on their data without relinquishing control and potentially withholding research irrelevant information from the researcher \cite{de2012trusted}. Implementation of this system requires more coordination than simply a plug-and-play kit, but partnering with a system that embraces this design allows users (research subjects) to set permissions for multiple studies, change permissions later, and revoke data access when desired. The Centers for Disease Control and Prevention (CDC) in the United States offers a similar setup for public health researchers requesting access to sensitive data through Remote Data Access Centers or by running submitted code queries \cite{cdc2015research}. A notable difference between these secure remote access options and the open PDS system design is the review of code run on sensitive data by federal specialists that review privacy implications and implement data use agreements that protect privacy and limit use to only approved research questions.

Developing an infrastructure that supports ethical and privacy-preserving research including data access requires additional investment. Even if the benefits to the developer are meaningful, the timescale for recouping on the investment may be a deterrence. However the presence of a free or low-cost platform that shifts the cost to the data user instead of the provider could make this a more viable option in the future. Additionally some robotics manufacturers, particularly those developing their robotic platforms for therapeutic applications, might recognize the benefit of external research using their platforms by validating or improving therapies offered. Data sharing platforms could be designed to give data sharers power to set data use agreements as a way to extend some control over how the data are used, which would further incentivize sharing by private companies to outside researchers.

\textbf{Anticipate New Knowledge and Unintended Consequences:} The rich and varied data collected by robots within intimate contexts like the home will require designers to carefully consider the acquisition of unintended knowledge and its consequences. Therapeutic robotics companies should anticipate detecting sensitive family matters encountered during the course of treatment or during acute times of crisis (e.g., end of life care) and have a response plan. This would not require a large financial investment from the robot manufacturer, but rather a development environment that encourages communication between a diverse team, relevant experts, and users about how the product might be used or integrated into daily life.

\section{Conclusion and Future Work}
\label{sec:conclusion}
These principles and recommendations are not intended to be comprehensive or definitive. Rather, it serves as a starting point for dialog between the robotics community and the privacy and research ethics communities so that the immense societal benefits of therapeutic robotics may be fully realized. By incorporating attention to relevant values into the information system design of therapeutic robotics early on in the commercialization of these technologies, we aim to improve the potential for ethical long-term research and access by a diverse set of researchers and practitioners. Whether commercial therapeutic robot manufacturers adopt our recommendations or not, we hope they consider these issues prior to releasing these powerful products broadly into the marketplace.  

Parts of this work remain speculative since many therapeutic robotic devices are only just now coming to the consumer market from research labs, so more analysis of how these devices are marketed and used will be required in the near future. Further, the law and policy landscape surrounding therapeutic robotic devices is continuing to evolve and will likely change and develop in the next few years. Research examining the treatment of human subjects (including their data) within the private sector is only beginning to emerge and requires further exploration into the current practices of private entities. Future work should examine which practices may be considered "generalizable research" and how norms or standards of practice for the consideration of human subjects should or should not be incorporated into private R\&D.  Economic incentives of private companies to consider privacy-preserving and ethical research practices will also likely be a large barrier to adoption and deserves further thought and analysis. 

\section{Acknowledgements}
\label{sec:acknowledgements}
This research was supported by the Hewlett Foundation through the UC Berkeley Center for Long-Term Cybersecurity (CLTC). The authors would also like to acknowledge the reviewers for their valuable feedback. The final publication is available at http://link.springer.com/article/10.1007/s12369-016-0362-y



\bibliographystyle{spmpsci}      
\bibliography{scm2016}   

\end{document}